# Detecting Molecular Hydrogen ($H_2$) Emission at Cosmic Dawn

Madisen Johnson[1], Blakesley Burkhart[1,2], Francesco D'Eugenio[3,4], Sandro Tacchella[3,4], Roberto Maiolino[3,4], Shmuel Bialy[5], Jacques Le Bourlot[6,7], Evelyne Roueff[6], Franck Le Petit[6], Emeric Bron[6], Hervé Abgrall[6], Erica Nelson[8], Shyam Menon[1,2], and Matthew E. Orr[1,2]

[1] Rutgers University, Department of Physics and Astronomy, Piscataway, NJ, USA
[2] Center for Computational Astrophysics, Flatiron Institute, 162 Fifth Avenue, New York, NY 10010, USA
[3] Kavli Institute for Cosmology, University of Cambridge, Madingley Road, Cambridge, CB3 0HA, UK
[4] Cavendish Laboratory, University of Cambridge, 19 JJ Thomson Avenue, Cambridge, CB3 0HE, UK
[5] Technion—Israel Institute of Technology, Haifa, Israel
[6] LUX, Observatoire de Paris, Université PSL, Sorbonne Université, CNRS, 92190 Meudon, France
[7] Université Paris Cité, Paris, France
[8] Department for Astrophysical and Planetary Science, University of Colorado, Boulder, CO 80309, USA

## Abstract

We present a theoretical framework for interpreting far-ultraviolet (FUV) fluorescent emission from molecular hydrogen ($H_2$) in high-redshift galaxies, motivated by the unique capabilities of the James Webb Space Telescope (JWST) to probe the rest-frame FUV at cosmic dawn. Using the Meudon photodissociation region code, we model the $H_2$ fluorescence spectrum under extreme interstellar medium (ISM) conditions in terms of high pressure ($10^{11}$ K cm$^{-3}$) and high radiation field ($10^6\,G_0$), combined with low metallicity ($Z = 0.1\,Z_\odot$) and high cosmic ionization rate ($\zeta = 10^{-14}$ s$^{-1}$), characteristic of early galaxies. As a case study, we apply this framework to stacked NIRSpec spectra from the JWST Advanced Deep Extragalactic Survey for galaxies at redshifts $z \geqslant 7$. The stacked spectrum exhibits emission features consistent in profile and wavelength with the predicted $H_2$ fluorescence lines, including a blueshift suggestive of an outflow of molecular gas. Although individual features remain below robust detection thresholds, this demonstration illustrates the feasibility of using FUV fluorescence modeling to guide and interpret JWST spectroscopy of the molecular ISM at high redshift. Our framework provides a foundation for future searches for molecular hydrogen emission and the study of galactic feedback processes in the early Universe.

## 1. Introduction

Molecular hydrogen ($H_2$), the most abundant molecule in the Universe, plays a crucial role in the baryonic life cycle throughout cosmic history. $H_2$ serves as a vital coolant during the early stages of the Universe (V. Bromm 2013; S. Bialy & A. Sternberg 2019; R. Klessen 2019; L. J. Tacconi et al. 2020; Y. Nakazato et al. 2022) and is often essential for star and planet formation processes (M. Chevance et al. 2023). Furthermore, $H_2$ is critical for the formation of complex molecules such as CO, OH, HCN, and $H_2O$, which cool dense gas (e.g., E. Herbst & W. Klemperer 1973; A. Sternberg & A. Dalgarno 1995; A. G. G. M. Tielens 2013; S. Bialy & A. Sternberg 2015). The inferred $H_2$ mass in galaxies strongly correlates with star formation rates and is a key diagnostic for star formation efficiency and gas depletion timescales (C. J. Lada et al. 2012; M. R. Krumholz 2014).

Direct detection of $H_2$ is challenging because it is a symmetrical molecule without a dipole moment. Therefore, only weak electric quadrupole emission may take place from its first excited rotational state, $J = 2$, which requires temperatures of $T = 511$ K, while most star-forming clouds have temperatures around $T \sim 10$ K. Consequently, $H_2$ is observable in warmer regions through infrared emissions, such as those captured by JWST MIRI-medium-resolution spectrometer at low redshift. However, detecting colder $H_2$ typically requires relying on less abundant proxies, such as CO, which emits brightly at low temperatures. The CO $J = 1 \rightarrow 0$ transition, with a critical temperature of 5.5 K, is often used to estimate $H_2$ masses using a conversion factor ($X_{CO}$; A. D. Bolatto et al. 2013; K. M. Sandstrom et al. 2013; C. Correia et al. 2014; C.-Y. Hu et al. 2022; S. Ganguly et al. 2024). However, variations in $X_{CO}$ due to metallicity, opacity, and cosmic-ray ionization rates introduce significant uncertainties, especially in extreme environments and at high redshifts (D. Narayanan et al. 2011; A. D. Bolatto et al. 2013; N. Imara & B. Burkhart 2016; H. H.-H. Chen et al. 2018; N. M. Pingel et al. 2018; B. A. L. Gaches et al. 2019; E. M. A. Borchert et al. 2022; D. Seifried et al. 2020).

Detecting CO at $z > 4$ is further complicated by its faintness and the observational preference for higher-J transitions. These transitions are favored because they are more easily excited in the warm, dense environments typical of high-redshift galaxies, have higher intrinsic brightness due to larger Einstein A coefficients, and are redshifted into Atacama Large Millimeter/submillimeter Array (ALMA) accessible bands at millimeter wavelengths. However, higher $J$ lines probe only the excited component of the molecular gas and require modeling to interpret, increasing additional uncertainties in $X_{CO}$. As a result, alternative tracers, such as [C II], [C I], and dust continuum emission, have been used to estimate the molecular gas mass at high redshifts (F. Walter et al. 2003; A. D. Bolatto et al. 2013; B. P. Venemans et al. 2017; L. J. Tacconi et al. 2020; Y. Ono et al. 2022; T. Hashimoto et al. 2023; G. C. Jones et al. 2024). In particular, M. Novak et al. (2019) reported stacked CO observations at $z = 7.5$ using ALMA.

Direct detection of $H_2$ is feasible through fluorescence and collisional excitation (e.g., shock-excited warm $H_2$ emitting in the IR). Relevant to this work, $H_2$ absorbs far-ultraviolet (FUV) photons in the Lyman–Werner band (912–1110 Å), exciting the molecule to electronic states ($B^1\Sigma_u^+$, $C^1\Pi_u$). Subsequent deexcitation produces FUV fluorescent emission lines, with approximately 85% of transitions leading to a vibrational-rotational cascade (T. L. Stephens & A. Dalgarno 1972; H. Abgrall et al. 2000). In addition, other excitation sources, such as cosmic rays, X-rays, and Ly$\alpha$ pumping, can also induce fluorescent emission (A. Sternberg 1988; S. Tiné et al. 1997; S. Bialy 2020; B. A. L. Gaches et al. 2022; M. Padovani et al. 2024).

As shown in A. Sternberg (1989), these FUV line intensities depend on gas density, UV radiation intensity, molecular formation rates, and dust absorption cross sections, while relative line intensities remain fairly insensitive to these parameters. However, these results are stated to be valid for $T \sim 500$ K and $n < 10^5$ cm$^{-3}$. Our high-redshift galaxies exceed these conditions, and future work will study the $H_2$ fluorescent emission lines in a range of conditions to investigate the relative line intensities in early Universe scenarios.

At low redshifts, space-based telescopes or stratospheric balloons are required for FUV $H_2$ observations due to atmospheric absorption. FUV fluorescent $H_2$ has been detected in nearby molecular clouds (such as the nearby Eos Cloud seen in B. Burkhart et al. 2025), superbubbles, and emission nebulae (K. France et al. 2004; K. France & S. R. McCandliss 2005; D. H. Lee et al. 2006; K. Ryu et al. 2006; Y.-S. Jo et al. 2011, 2015, 2017; T.-H. Lim et al. 2015). The origin of the fluorescence is not fully understood, as it might arise from a variety of processes. Proposed missions like Hyperion and Eos aim to further study the $H_2$ fluorescence in the Milky Way (E. T. Hamden et al. 2022, 2024).

## 2. Modeling High-redshift $H_2$

### 2.1. Meudon Photodissociation Region Model

Detecting electronic excited states of $H_2$ requires the observation of several key bright features of the $H_2$ emission spectrum, which are a combination of the discrete and oscillating continuum Lyman and Werner band systems of molecular hydrogen as computed by H. Abgrall et al. (2000). $H_2$ continuum lies between 135 and 165 nm and results from emission from electronically excited states toward the dissociating continuum, which leads to the destruction of the molecule (T. P. Stecher & D. A. Williams 1967). The continuum emission was first observed in the laboratory and was identified by A. Dalgarno & T. L. Stephens (1970) and A. Dalgarno et al. (1970).

We use the Meudon photodissociation region (PDR) code (F. Le Petit et al. 2006) to compute a model spectrum of lines and continuum for electronically excited $H_2$ (F. Le Petit et al. 2006; J. R. Goicoechea & J. Le Bourlot 2007; M. Gonzalez Garcia et al. 2008; J. Le Bourlot et al. 2012; E. Bron 2014; E. Bron et al. 2014, 2016). For our analysis, we focus on high-density and high-pressure conditions to capture the extreme conditions of early Universe observations. We will then use this model to compare with high-redshift galaxy spectra.

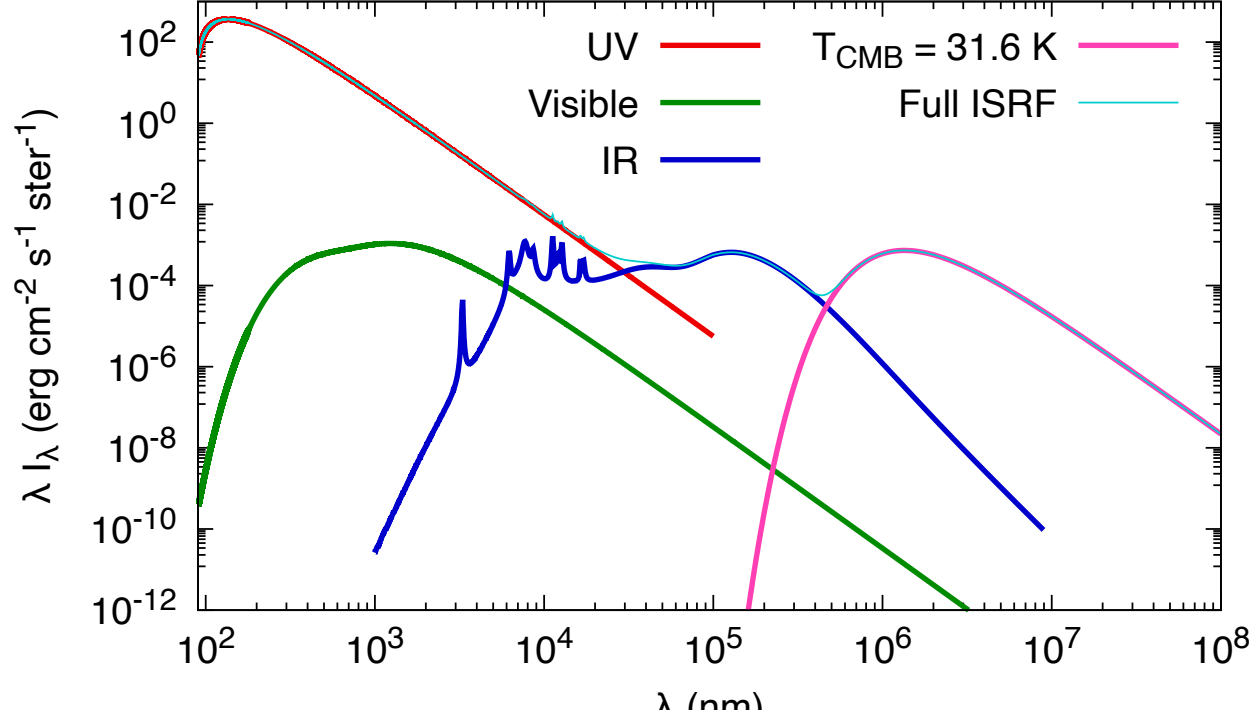

**Figure 1.** High-redshift Meudon model radiation field, including the cosmic microwave background (pink), IR (blue), visible (green), and the B. T. Draine (1978) UV component, which is scaled to provide an incident $G_0 = 10^6$ at the cloud's edge (red).

**Table 1**
Model Input Parameters

| Parameter | Value |
|---|---|
| $A_{V,\max}$ | 10 |
| $G_0$ | $10^6$ |
| $\zeta$(s$^{-1}$) | $10^{-14}$ |
| $P/k_B$(K cm$^{-3}$) | $10^{11}$ |
| $Z$ | 0.1 |
| D/G | $10^{-4}$ |

**Note.** Maximum $A_V$ (cloud size), strength of UV component of the ISRF ($G_0$), cosmic-ray ionization rate ($\zeta$), pressure ($P$), metallicity ($Z$) relative to local Galactic values, and dust-to-gas ratio (D/G), which is approximately a factor of 100 lower than the local Galactic value.

The Meudon PDR code considers a stationary plane-parallel slab of gas and dust illuminated by a radiation field that extends from ultraviolet to microwave (seen in Figure 1; we note that the ultraviolet component is extrapolated by a power law up to $\lambda = 10^5$ nm), coming from one or both sides (where the two intensities can be different). The code then solves radiative transfer over the full wavelength range at each point in the cloud, taking into account absorption in the continuum by dust and gas and in the discrete transitions of H and $H_2$ and possibly CO and their isotopes. The thermal balance is then computed, taking into account heating processes such as the photoelectric effect on dust, ultraviolet pumping, and radiative cascades of $H_2$, followed by collisional deexcitation, exothermic chemistry, cosmic rays, etc., and cooling from line emission of the abundant ions, atoms, and/or molecules and coupling between gas and dust. Chemistry computation is strongly coupled to radiative transfer and thermal balance since reaction rates depend on the gas and grain temperatures, the UV flux at each position, the charge of grains, etc. This chemistry computation is then used to calculate the abundance of each species at each point. This allows the program to calculate column densities, emissivities, and intensities.

For the model used in this analysis, we use the input parameters from Table 1. These values were chosen as typical of high-redshift, early-Universe conditions. In particular, the extremely high pressure ($10^{11}$ K cm$^{-3}$) and the high radiation field of $G_0 = 10^6$, where $G_0$ is defined as the strength of the FUV field normalized to the average interstellar radiation field in the solar neighborhood prescribed by B. T. Draine (1978), are well above the conditions normally found within our Galaxy. This includes high-density and high-pressure conditions that are in line with direct measurements of high electron densities (Y. Isobe et al. 2023) and high temperatures, even in star-forming galaxies (e.g., M. Curti et al. 2023; I. H. Laseter et al. 2024; D. Schaerer et al. 2024), but should be considered as typical orders of magnitude. The link between metallicity and the dust-to-gas mass ratio is taken from F. Galliano et al. (2021, see their Figure 14). The public version of the code[9] has been updated to include the higher excited electronic states of $H_2$ ($B'\,^1\Sigma_u^+$ and $D\,^1\Pi_u$) in the absorption and fluorescence processes. Fluorescence includes discrete line emission within and to the $X\,^1\Sigma_g^+$ ground electronic state and continuum radiation above the dissociation limit of $H_2$.

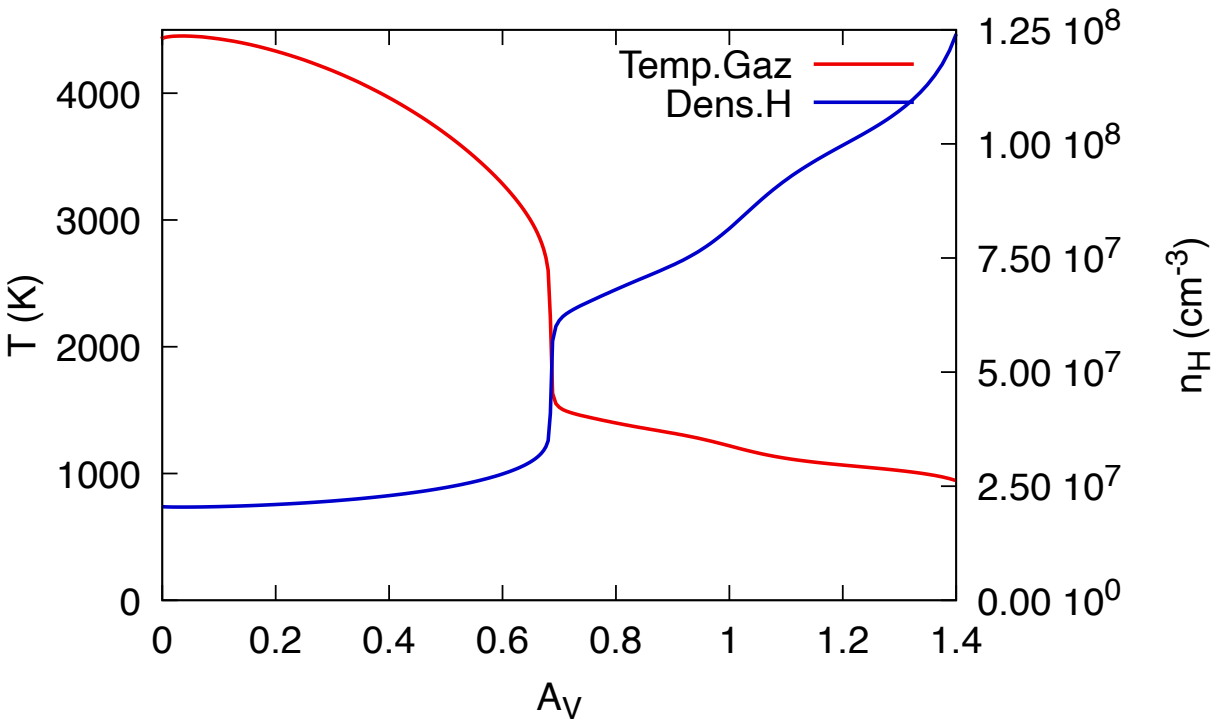

**Figure 2.** Temperature (red line) and density (blue line) profiles of the illuminated cloud in our Meudon PDR model. The profiles showcase two distinct regions: a hot and diffuse outer envelope, which transitions to a cooler and denser region deeper into the cloud.

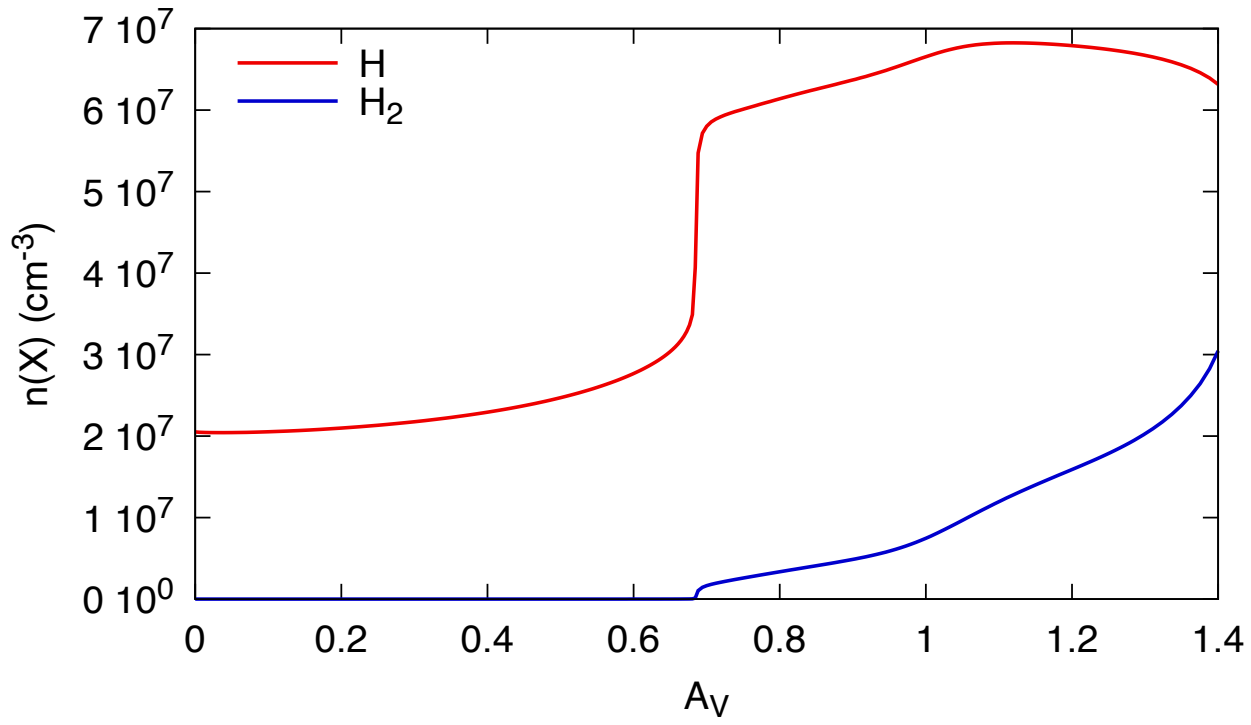

**Figure 3.** H and $H_2$ profiles of the illuminated cloud and its two distinct regions. The outer envelope composed of fully atomic gas, and the inner region contains a significant amount of warm $H_2$ gas.

Figure 2 shows the temperature and density structure at the edge of the PDR. Beyond $A_V = 1.4$, the fluid undergoes a strong thermal instability that brings the deeper part of the cloud to dark cloud conditions that are irrelevant to the present study, since the electronic states $H_2$ cannot be populated anymore. The outer part exhibits two regions:

1. From $A_V = 0$ to 0.68: the gas is fully atomic and very hot ($T > 4000$ K).
2. From $A_V = 0.68$ to 1.45: there is a significant amount of $H_2$ within a mostly warm atomic gas (where $T \sim 1000$ K).

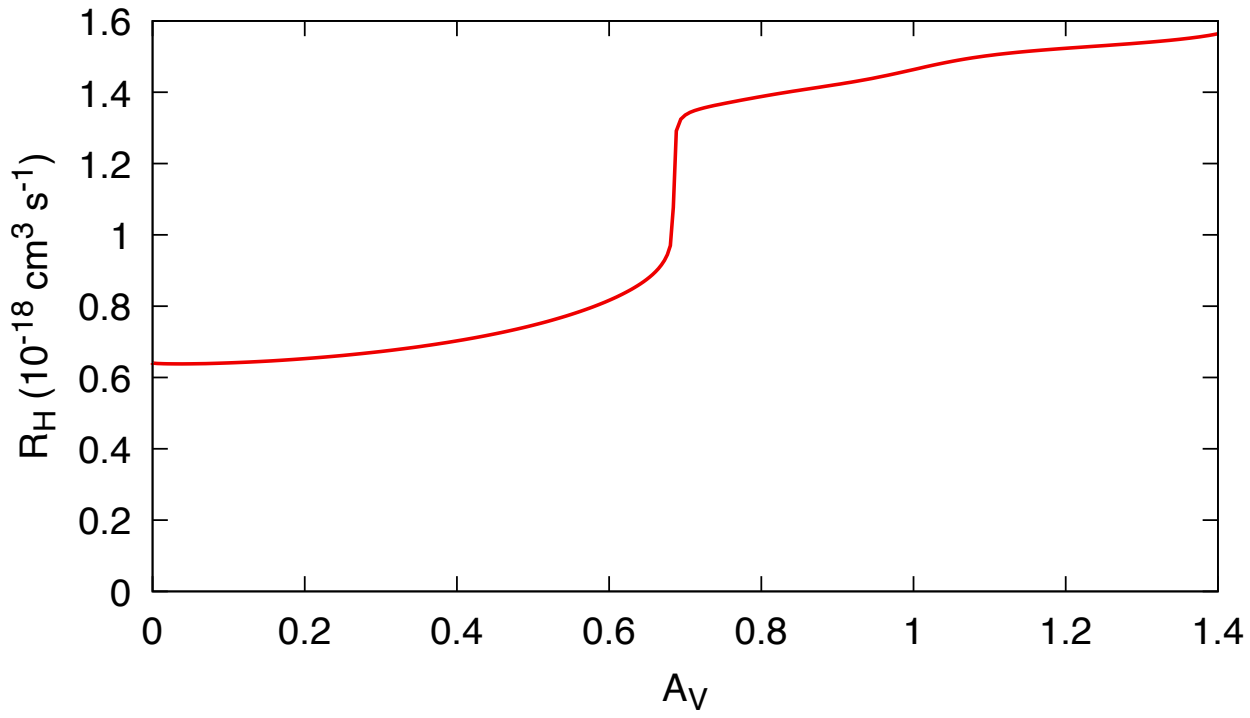

**Figure 4.** Effective formation rate coefficient of $H_2$.

The transition from region 1 to region 2 marks the point where $H_2$ self-shielding is efficient enough to allow the survival of the molecule (A. Sternberg et al. 2014), which can be seen in the H and $H_2$ profiles in Figure 3. Before that, $H_2$ destruction is dominated by photodissociation. The strong drop in temperature in a constant-pressure cloud leads to a large increase in H and the emergence of $H_2$ in region 2. This second region provides a significant amount of warm molecular hydrogen and is still bathed in a significant amount of UV radiation.

Beyond $A_V = 0.68$, destruction is dominated by very active chemistry, driven by warm and dense conditions that allow endothermic reactions or reactions with activation barriers to occur due to vibrationally excited $H_2$. The main reactions are as follows:

$$H_2 + O \longrightarrow H + OH$$
$$H_2 + OH \longrightarrow H + H_2O$$
$$H + OH \longrightarrow H_2 + O$$
$$H_2O + h\nu \longrightarrow H_2 + O$$
$$H + H_2O \longrightarrow H_2 + OH.$$

Reaction with $C^+$, following

$$C^+ + H_2 \rightleftharpoons CH^+ + H$$

is also very fast, but the forward and backward rates almost cancel each other out.

Inspection of individual formation and destruction rates shows that these reactions lead to the net destruction of molecular hydrogen. Other reactions have a marginal impact and, in particular, formation through $H^-$ is completely negligible, which is due to its very high photodetachment rate.

The formation of $H_2$ occurs through the Eley–Rideal mechanism on warm grains (F. Le Petit et al. 2006; J. R. Goicoechea & J. Le Bourlot 2007; M. Gonzalez Garcia et al. 2008; J. Le Bourlot et al. 2012; E. Bron 2014; E. Bron et al. 2014, 2016). Due to the fast evaporation of physisorbed hydrogen atoms from hot grains, the Langmuir–Hinshelwood mechanism is negligible here. The abundance of chemisorbed H on the surface of the dust grains is calculated as a function of grain size. Then, the direct impact of the fast gas-phase hydrogen atoms allows the ability to overcome the reaction barrier. The resulting rate coefficient is not constant and cannot be summarized by a single simple formula. Figure 4 shows $R_H$, calculated as

$$R_H = \left.\frac{dn(H_2)}{dt}\right|_{Form} \times \frac{1}{n(H)\, n_H}, \quad (1)$$

[9] Available at https://pdr.obspm.fr.

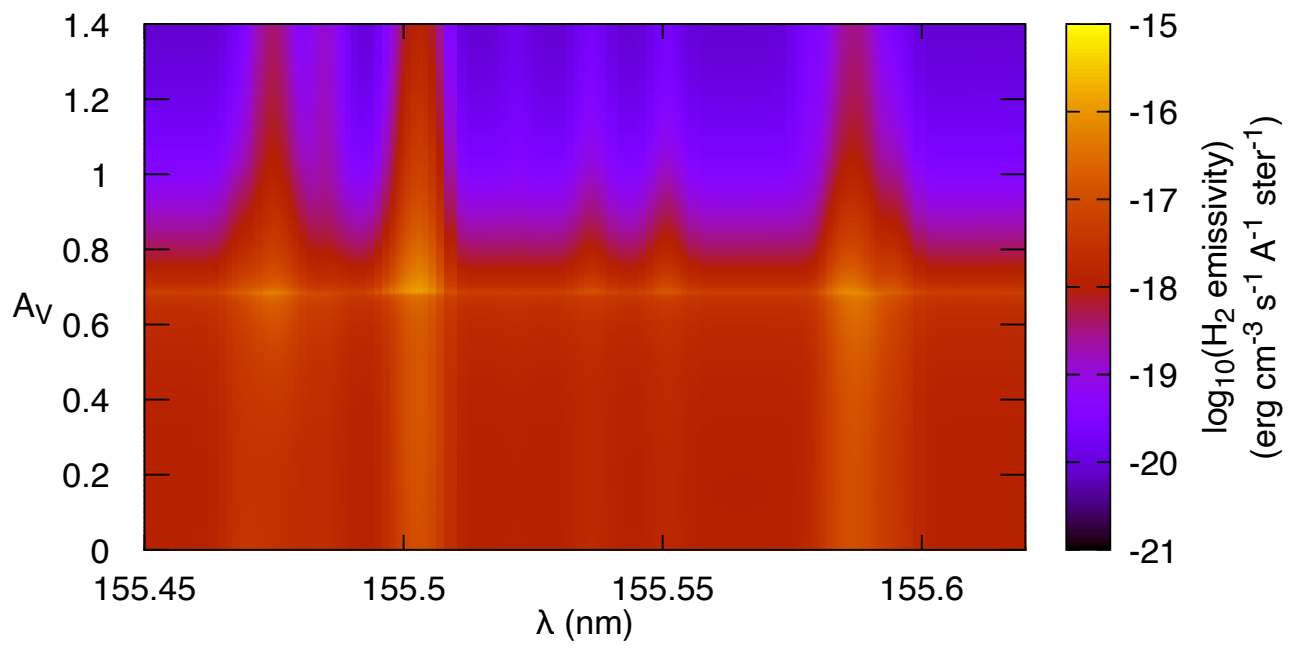

**Figure 5.** $H_2$ emissivity spatial repartition in a small range of wavelength (measured in the local rest frame). The model predicts a strong and rather uniform emissivity leading to the continuum part of the fluorescence spectrum in region 1 (lower part of the map). In region 2 (upper part of the map), only emission lines remain.

where $n(\mathrm{H})$ and $n(\mathrm{H_2})$ are the abundances of atomic and molecular hydrogen in $\mathrm{cm^{-3}}$, and $n_\mathrm{H}$ is the total hydrogen density in $\mathrm{cm^{-3}}$. The increase in efficiency at the $\mathrm{H/H_2}$ transition is due to the variation in the chemisorption sticking coefficient with temperature, as detailed in J. Le Bourlot et al. (2012, App. C.2).

$H_2$ emission from regions 1 and 2 is qualitatively different:

1. In region 1, each $H_2$ molecule is efficiently destroyed as soon as it is formed. This leads to a strong continuum emission even if the resulting $H_2$ abundance remains very low.
2. In region 2, there is enough $H_2$ for self-shielding to become efficient, and excitation effects to emerge. Relaxation to $v = 0$ allows emission in lines to participate in the total flux, and continuum emission becomes negligible.

This is visible in Figure 5, which shows $H_2$ contribution to emissivity[10] in a small range of wavelength as a function of depth into the cloud. In region 1 (lower part of the map), we find a strong and rather uniform emissivity that leads to the continuum part of the fluorescence spectrum. In region 2 (upper part of the map), only emission in lines remains, with lower $J$ lines being stronger and going deeper into the cloud than high $J$ lines.

In $H_2$ photodissociation, the number of emitted photons in the continuum is equal to the number of absorbed photons leading to dissociation. Thus, we can evaluate the total integrated emissivity of the continuum contribution to fluorescence. Integrating for $H_2$, from the edge of the cloud to the depth of the emitting region $L_\mathrm{max}$, the product of the photodissociation rate $k_D$ times the local abundance $n_{\mathrm{H_2}}$ gives the total number of photons emitted in continuum fluorescence. If we attribute to these photons a typical wavelength of $\lambda_0 = 150$ nm, then the integrated intensity of the continuum is

$$I_\mathrm{cont} = \frac{h\,c}{\lambda_0}\,\frac{1}{4\pi}\int_0^{L_\mathrm{max}} k_D(s)\,n_{\mathrm{H_2}}(L)\,dL. \qquad (2)$$

This gives $I_\mathrm{cont} = 0.2\ \mathrm{erg\ cm^{-2}\ s^{-1}\ sr^{-1}}$. Note that this is about 10% of the total fluorescence integrated intensity, as, in the mean, 1 photon every 10 leads to dissociation and not to line emission toward the observer. A typical fluorescence spectrum is shown in Figure 6, convolved at a resolution $R = 1000$ to best compare with JWST spectra. With such a low resolution, narrow or weak emission features coming from electronic states above the Lyman and Werner systems are washed out.

[10] The emissivity is in $\mathrm{erg\ cm^{-3}\ s^{-1}\ \AA^{-1}\ sr^{-1}}$, that is, a specific intensity per unit length as found in the radiative transfer equation.

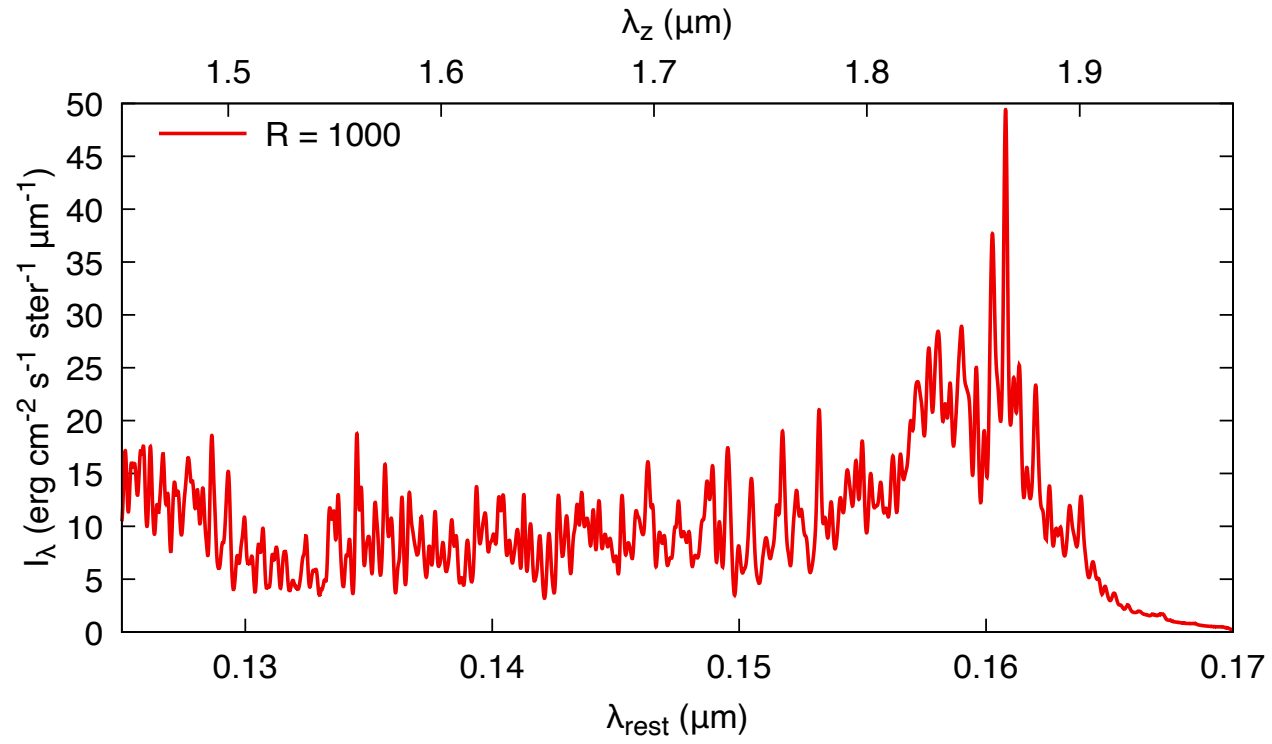

**Figure 6.** Fluorescence spectrum of $H_2$ at a resolution of $R = 1000$ produced by our Meudon PDR model. Using this spectrum, we can then identify the strongest features and compare to observations. The upper wavelength range assumes a redshift $z = 10.6$, which is the highest redshift source in our sample.

## 3. Application to JADES Galaxies

### 3.1. Data and Methodology

We utilize data from the James Webb Space Telescope (JWST) Advanced Deep Extragalactic Survey (JADES), which provides extensive infrared imaging and spectroscopic observations in the GOODS-South (GOODS-S) and GOODS-North (GOODS-N) deep fields (D. J. Eisenstein et al. 2023a, 2023b; A. J. Bunker et al. 2023; F. D'Eugenio et al. 2025). This survey is designed to investigate the evolution of galaxies from the epoch of reionization to cosmic noon. JADES draws on approximately 770 hr of Cycle 1 guaranteed observation time from the Near-Infrared Camera and the Near-Infrared Spectrograph (NIRSpec), offering a rich dataset ideal for our analysis. All data used in this study are publicly available through the JADES website and the Mikulski Archive for Space Telescopes (M. Rieke et al. 2023).

Our analysis specifically focuses on the publicly available JADES dataset, which includes medium-depth NIRSpec multiobject spectroscopy. This spectroscopy covers the wavelength range of 0.6–5.3 $\mu$m using five dispersers. The dataset builds on the extensive legacy of the GOODS-S and GOODS-N fields, developed over the past 30 yr, now further enhanced by the capabilities of JWST.

The NIRSpec spectroscopic data were reduced using the standard JWST pipeline. The reduction process included detector-level corrections (e.g., bias subtraction and flat-fielding), wavelength and flux calibration, and the extraction of one-dimensional spectra. Additional postprocessing steps were applied to refine background subtraction and mitigate contamination from overlapping spectra in crowded regions. A detailed description of the data reduction procedure is provided in D. J. Eisenstein et al. (2023a, 2023b).

To search for $H_2$ emission with JWST, we selected all galaxies above a redshift of 7 with the redshift measurement quality flagged with "A." We analyzed 59 galaxies (6 in GOODS-N and 53 in GOODS-S) from the JADES catalog with $z = 7 - 10.6$, yielding a mean redshift of 7.92. Below a redshift of 9.4, redshifts are determined from strong rest-frame optical emission lines, such as [O III] $\lambda\lambda$4959,5007, which yield redshift uncertainties of approximately one-tenth of a pixel (D. J. Eisenstein et al. 2023a). However, at redshifts higher than this, they are almost always determined from low equivalent width rest-frame UV lines or from the Ly$\alpha$ break. Due to the uncertainty of this second method, we have therefore included only sources with redshift measurements flagged as the best quality.

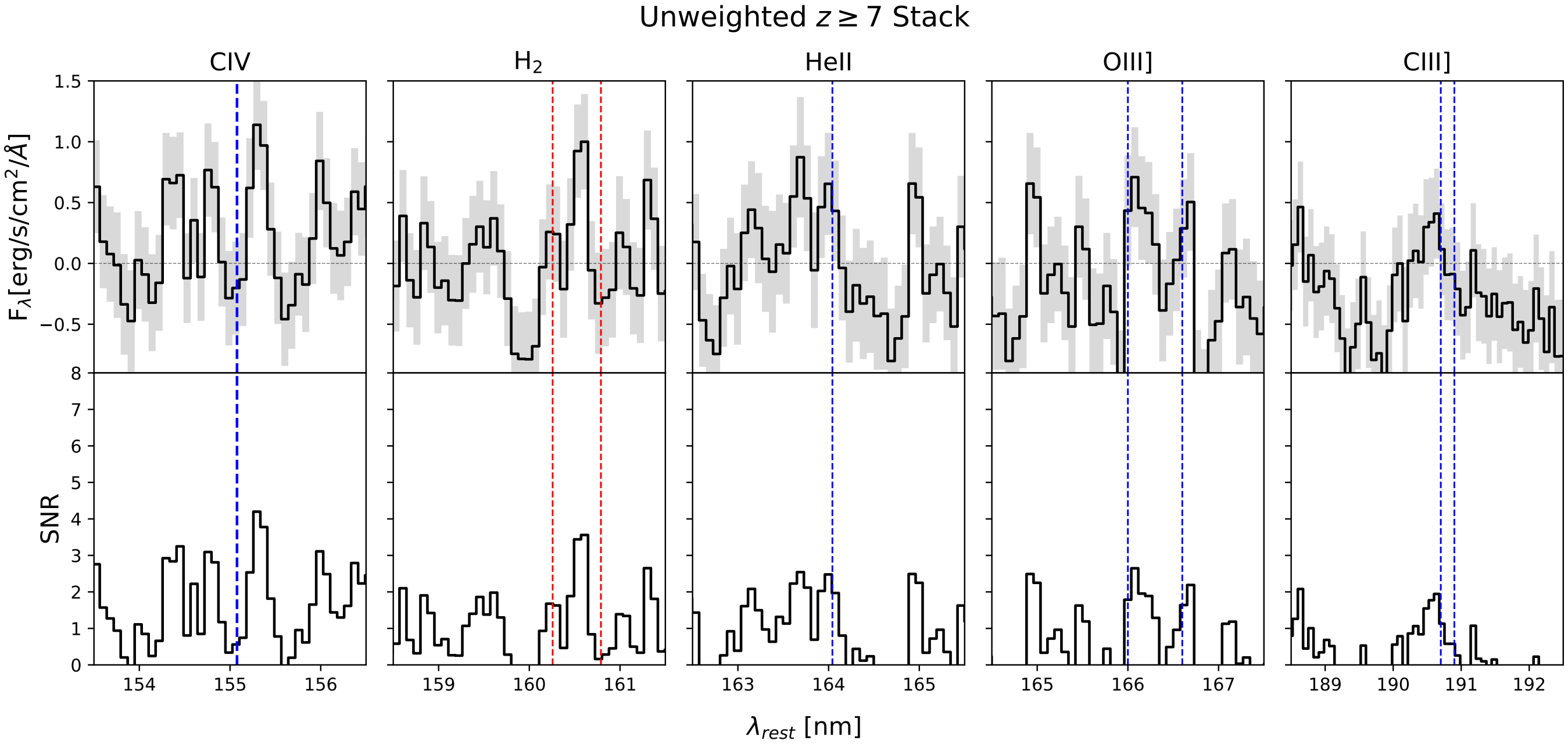

**Figure 7.** Continuum-subtracted stacked spectrum (top row) and the signal-to-noise ratio (bottom row) of all JWST JADES galaxies with redshifts $z \geqslant 7$, investigating the tentative detections of C IV, $H_2$ Fluorescence, He II, O III], and C III]. These potential detections hint at the presence of multiphase outflows at cosmic dawn through slightly shifted emission peaks and P Cygni–like profiles.

### 3.2. JADES $z \geqslant 7$ Spectral Stack

The emission features seen in Figure 6 indicate that the brightest FUV $H_2$ line feature in the JWST band would be observed at $\lambda_{\rm rest} = 0.16\,\mu$m. It is important to note that these are emission features that are composed of multiple emission lines.

To attempt to detect molecular hydrogen fluorescence signatures, we began by stacking spectra from numerous high-redshift galaxies to enhance faint signals. This process involved shifting all galaxies to their rest frames and regridding the spectra to match the resolution of the JWST instruments. Each spectrum was normalized using its maximum intensity within the UV range (140–170 nm) before stacking. This normalization is used to ensure that the signal from each galaxy contributes equally to the overall stacked signal. This allows us to understand the overall properties of the sample instead of the properties of the brightest galaxies.

Stacking was performed by summing the flux values at each wavelength due to the inhomogeneous properties of the galaxies within our sample. For example, some studies suggested that high-redshift galaxies undergo rapid star formation quenching, via removal of most or all of the galaxy's interstellar medium (ISM; e.g., A. Dressler et al. 2023; V. Strait et al. 2023; A. Ferrara 2024; T. J. Looser et al. 2024; W. M. Baker et al. 2025; W. McClymont et al. 2025), which would mean that the brightest galaxies (in terms of stellar UV continuum) may not have the highest EW emission lines. However, we have also performed an inverse-variance stacking, which is presented and discussed in the Appendix. We then subtracted the continuum of the spectral stack by fitting a power-law slope while excluding our features of interest. The resulting stacked spectrum was then compared to our high-redshift $H_2$ Meudon model, as described in the previous section. We note that due to the Gunn–Peterson trough at these redshifts, we must limit our search to bright features redward of the Ly$\alpha$ break.

The stack spectra produced by this set of galaxies are shown in Figure 7 (top row). In addition to the stacked spectra, we plotted the signal-to-noise ratio (SNR) using a moving boxcar approach (bottom row). We first note a prominent P Cygni–like profile (redshifted emission peak and small blueshifted absorption trough) of the C IV emission line. This profile is indicative of outflow material and corresponds to an average outflow velocity range of 600–1000 km s$^{-1}$. Ionized atomic outflows have been shown to be common at redshifts around cosmic dawn, and these spectra provide additional evidence for the ubiquity of galactic outflows in the early Universe.

We then compare our tentative $H_2$ features with the Meudon PDR model in Figure 8. To achieve the best fit between the model and the stacked data, the $H_2$ model was blueshifted by 0.18 nm, corresponding to an outflow velocity of approximately 340 km s$^{-1}$, which is in agreement with the observed outflow predicted by the P Cygni–like C IV line. Additionally, we note that this feature also displays a P Cygni–like profile with an absorption trough that is even deeper than that of the C IV.

This blueshift and profile indicate that the observed $H_2$ emission may predominantly be from gas moving toward the observer, consistent with an outflow scenario in which the receding component is obscured, likely due to dust extinction (J. H. Matthews et al. 2023). FUV fluorescence lines of $H_2$ are particularly susceptible to dust attenuation, which may suppress signals from the receding side of the outflow. While stacking enhances faint signals, some galaxies in the sample may not significantly contribute to the outflow signal because of orientation effects, such as an outflow perpendicular to the line of sight, where the projected radial velocity is minimal.

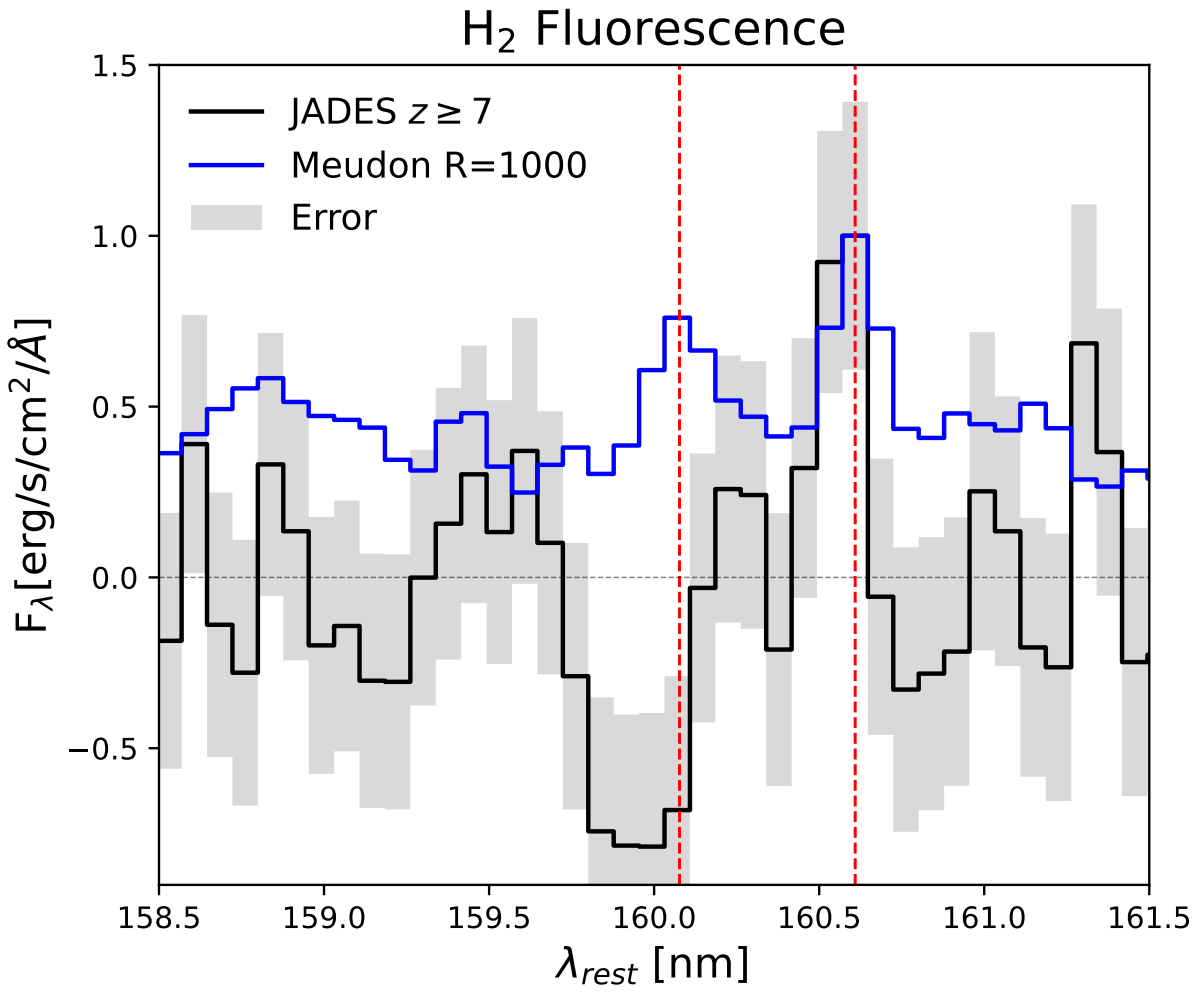

**Figure 8.** Comparison of the potential stacked $H_2$ fluorescent emission feature with our high-redshift Meudon PDR model (blue line) and predicted features (red vertical lines), which have now been shifted by 0.18 nm or ∼340 km s$^{-1}$ to better match the observed spectrum. This displays the unique profile of the potential detection, which could indicate the presence of molecular outflows.

These limitations underscore the importance of follow-up analyses with individual spectra to disentangle contributions from different galaxies and confirm the presence of outflows. Previous studies, such as those investigating [C II] outflows (F. Stanley et al. 2019), have used stacking to identify outflows, supporting the validity of this approach.

Additional spectral features, such as He II, O III] $\lambda$1660, 1666, and C III] $\lambda$1907, 1909, are also investigated. These features can further constrain the average conditions of these galaxies and indicate an overall high-density and high-temperature environment (A. Feltre et al. 2016; L. Pentericci et al. 2018; O. Le Fèvre et al. 2019; A. Grazian et al. 2020). These properties are characteristic of galaxies with high star formation rates or active galactic nuclei (AGN). For example, the ratio of the C III] doublet can be utilized to diagnose the density of galactic environments. However, because of the inability to resolve both doublet peaks, it is difficult to make any claims about the general properties of the galaxy population. Similarly, the O III] feature, which can be used to diagnose temperature conditions within the galaxy, is also not fully resolved. In future work, we hope to further look into sources within the spectrum stack and use these diagnostics to determine the properties of galaxies and how they potentially affect fluorescent molecular hydrogen in the early Universe.

When examining the SNR, as shown in Table 2, we find that C IV and He II are detected with significances of SNR = 4.20 and SNR = 2.54, respectively. The most prominent feature of the $H_2$ fluorescent spectrum (emitted at 160.8 nm but observed at 160.61 nm) has a comparable SNR (SNR ∼ 3.56), while the next brightest $H_2$ feature (emitted at 160.3 nm but observed at 160.22 nm) has an SNR = 1.68. The low SNR of the second feature emphasizes the critical importance of follow-up observations, which may allow us to detect these features with greater certainty.

Although the intensities of UV fluorescent lines are sensitive to gas density, incident UV radiation intensity, molecular formation rates, and grain UV absorption cross section, the ratios of the lines are weakly sensitive to the spectral shape of

**Table 2**
Tentatively Detected Emission Lines, Their Measured Emission Wavelength, and Their SNR for the Stacking of JADES Galaxies With $z \geqslant 7$

| Emission Line/Feature | $\lambda_{obs}$ (Å) | SNR |
|---|---|---|
| C IV $\lambda\lambda$1548,1550 | 1552.96 | 4.20 |
| $H_2$ $\lambda\lambda$1603,1608 | 1602.24, 1606.09 | 1.68, 3.56 |
| He II $\lambda$1640 | 1636.89 | 2.54 |
| O III] $\lambda\lambda$1660,1666 | 1660.76, 1666.92 | 2.64, 2.19 |
| C III] $\lambda\lambda$1907,1909 | 1906.39, … | 1.94, … |

**Note.** The $H_2$ features cannot be attributed to a specific line but are instead a blend of multiple emission lines.

the incident UV field (A. Sternberg 1989). For example, in our high-redshift model, the $H_2$ fluorescence features scale with pressure, becoming stronger at $P/k_B = 10^{11}$ K cm$^{-3}$, weaker at $P/k_B = 10^9$ K cm$^{-3}$, and almost invisible at $P/k_B = 10^7$ K cm$^{-3}$. These trends, which can be seen in Figure 1 of A. Sternberg (1989), reflect the value of $G_0/n_H$ in the H/$H_2$ formation layer, where most of the emission originates, predicting high-pressure environments in these galaxies.

## 4. Implications and Future Work

This paper presents the first direct tentative detection of the $H_2$ fluorescent emission features at high redshifts. This suggests that JWST can be a powerful probe of molecular hydrogen gas in galaxies during the epoch of reionization. Our findings underscore the potential capability of JWST NIRSpec in detecting molecular hydrogen gas in emission at ultrahigh redshifts. The rest frame wavelength range of $H_2$ lines spans 91.1–164.4 nm (A. Sternberg 1989), while NIRSpec covers a wavelength range of 0.6–5 $\mu$m. This enables the possible characterization of warm fluorescent $H_2$ emission features in galaxies across a redshift range from $z = 5.6$ to beyond $z = 15$. Given the faintness of CO emission at such redshifts, our results offer a novel method for directly probing the molecular ISM in galaxies that lie beyond ALMA's observational capabilities. For $z < 7.5$, these observations complement studies of cold gas traced by dust, [C II], and CO.

The $H_2$ FUV emission lines serve as powerful probes to constrain the physical properties of the ISM in galaxies and to identify excitation sources. As discussed in G. B. Field et al. (1966), T. P. Stecher & D. A. Williams (1967), A. Dalgarno et al. (1970), and A. Sternberg (1989), when a molecular or partially molecular neutral gas is irradiated by FUV radiation, $H_2$ molecules are excited to electronic states. These states rapidly decay back to the ground electronic state, producing FUV line emission. The intensities of these emission lines are directly proportional to the flux of the incident FUV radiation. Consequently, the $H_2$ FUV lines can be used to constrain the interstellar FUV radiation field in galaxies. Since the FUV radiation is predominantly produced by massive, short-lived stars, these measurements can be further connected to constrain the population of massive stars. For a known initial mass function, this enables estimates of the star formation rate.

Furthermore, measurement of the intensity of the $H_2$ FUV line therefore allows one to determine the integrated $H_2$ photodissociation rate (B. Burkhart et al. 2024; S. Bialy et al. 2025). Comparing this quantity with other timescales, such as the Hubble time, the star formation timescale, or the dynamical timescale, can offer valuable insights into the evolution of the galactic ISM, the bottlenecks in its evolution, and the processes governing the buildup of stellar mass.

## 5. Conclusions

In this work, we develop and present a theoretical framework for modeling FUV fluorescent emission from molecular hydrogen ($H_2$) in high-redshift galaxies, with the goal of enabling future analyses of JWST spectroscopy during the epoch of reionization and beyond. Using the Meudon photodissociation region code, we simulate $H_2$ emission spectra under ISM conditions representative of the early Universe, including high radiation fields, low metallicities, and extreme pressures. These models predict characteristic line features in the 155–190 nm rest frame wavelength range that are largely insensitive to local microphysical parameters, offering a robust set of diagnostics to trace warm molecular gas.

To demonstrate the utility of this framework, we apply our models to a stacked sample of JWST/NIRSpec spectra from the JADES survey, focusing on galaxies with redshifts $z \geqslant 7$. The resulting stacked spectra show spectral features that are qualitatively consistent with our model predictions, including emission line profiles and modest blueshifts indicative of outflowing gas. Although the SNR and relative line intensities of these features are not yet sufficient for definitive detection, the agreement in spectral morphology highlights the promise of this approach.

Our findings support the use of FUV $H_2$ fluorescence as a complementary tool to CO and [C II] to investigate the molecular phase of ISM in the early Universe. The modeling framework presented here can be applied to both individual and stacked spectra, providing physical constraints on molecular gas properties, radiation field strength, and outflow kinematics in galaxies at $z \geqslant 6$.

In future work, we also aim to perform more detailed modeling (e.g., high-density effects), further analyze other emission lines (e.g., [Fe II], He II, O III], C III], and Ly$\alpha$ pumping), and the contribution of $H_2$ to the UV continuum, to better constrain the physical properties of the gas within multiphase outflows, such as average radiation field strength and mass loading factors. As JWST continues to expand its spectroscopic reach and as future UV missions such as Hyperion/Eos are developed, this methodology offers a path forward for robust detection and interpretation of molecular hydrogen across cosmic time.

## Acknowledgments

M.J. and B.B. thank Dr. Amiel Sternberg for the many discussions on $H_2$ fluorescence. B.B. acknowledges support from NSF grant AST–2009679 and NASA grant No. 80NSSC20K0500. B.B. is grateful for the generous support by the David and Lucile Packard Foundation and the Alfred P. Sloan Foundation. B.B. thanks the Center for Computational Astrophysics (CCA) of the Flatiron Institute and the Mathematics and Physical Sciences (MPS) division of the Simons Foundation for support. The Flatiron Institute is supported by the Simons Foundation. S.B. acknowledges support from the ISF grant No. 2071540, the GIF grant No. I–1568–303.7/2024, the NSF-BSF grant No. 2023761, and the Alon Fellowship prize for junior faculty. The Meudon group acknowledges the support of the Thematic Action "Physique et Chimie du Milieu Interstellaire" (PCMI) of INSU Programme National "Astro," with contributions from CNRS Physique & CNRS Chimie, CEA, and CNES. F.D. and R.M. acknowledge the support of the Science and Technology Facilities Council (STFC), by the ERC through the Advanced Grant 695671 "QUENCH", and by the UKRI Frontier Research grant RISEandFALL. R.M. also acknowledges funding from a research professorship from the Royal Society.

## Appendix

As previously mentioned in Section 3, assessing the inverse-variance weighting should increase the SNR of any detections. In practice, this works well only for homogeneous data. Our sample, in contrast, spans a broad range of galaxy properties, so it is not clear to us whether inverse-variance weighting is the best choice. For example, some studies suggested that high-redshift galaxies undergo rapid star formation quenching, via removal of most or all of the galaxy ISM (e.g., A. Dressler et al. 2023; V. Strait et al. 2023; A. Ferrara 2024; T. J. Looser et al. 2024; W. M. Baker et al. 2025; W. McClymont et al. 2025), which would mean that the brightest galaxies (in terms of stellar UV continuum) may not have the highest EW emission lines.

For this reason, we present both stacking approaches, with the inverse-variance weighting reported as follows:

The stack spectra produced by this set of galaxies are shown in Figure 9 (top row). In addition to the stacked spectra, we plotted the SNR using a moving boxcar approach (bottom row).

We first note a significant change in the C IV line profile. In contrast to the unweighted stacking results, the profile is now unclear, and there seems to be no convincing detection of the presence of C IV.

We then compare our tentative $H_2$ feature with the Meudon PDR model in Figure 10. To achieve the best fit between the model and the stacked data, the $H_2$ model was blueshifted by 0.18 nm, corresponding to an outflow velocity of approximately 340 km s$^{-1}$, which is in agreement with the observed outflow predicted by the P Cygni–like C IV line. Additionally, we note that this feature also displays a slight P Cygni–like profile. The profile of the $H_2$ features is similar to that seen in the unweighted regime.

Additional spectral features, such as He II, O III] $\lambda$1660, 1666, and C III] $\lambda$1907, 1909, are also investigated. These features can further constrain the average conditions of these galaxies and indicate an overall high-density environment (A. Feltre et al. 2016; L. Pentericci et al. 2018; O. Le Fèvre et al. 2019; A. Grazian et al. 2020). These properties are characteristic of galaxies with high star formation rates or AGN. Similarly to the previous results, the C III] lines are not resolved and therefore their ratios cannot be used to diagnose the environment. However, when looking at the O III] doublet, we see that the emission lines have both increased in their SNR. Unfortunately, we see that O III] 1660,1666 is approximately 2, which is greater than the theoretical value that can be obtained. This may be due to the stack containing a variety of galaxies with different properties. Despite the lack of a promising detection, it still provides an interesting feature to inspect for future studies to understand the environmental properties of galaxies where we may observe $H_2$.

When examining the SNR as shown in Table 3, we find that C IV and He II are detected with significances of SNR = 4.28

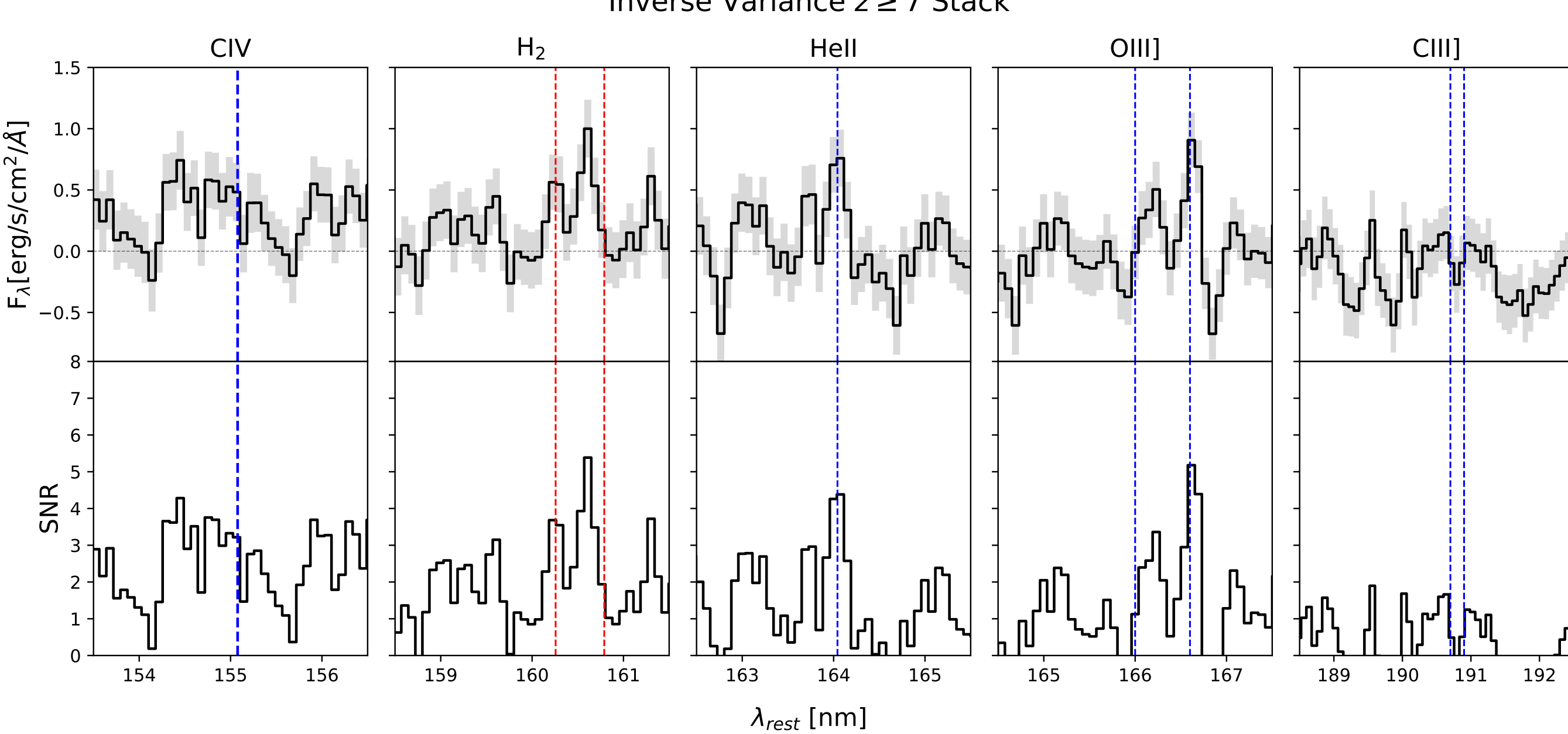

**Figure 9.** Inverse-variance stacked spectrum (top row) and the SNR (bottom row) of all JWST JADES galaxies with redshifts $z \geqslant 7$, investigating the tentative detections of C IV, $H_2$ Fluorescence, He II, O III], and C III].

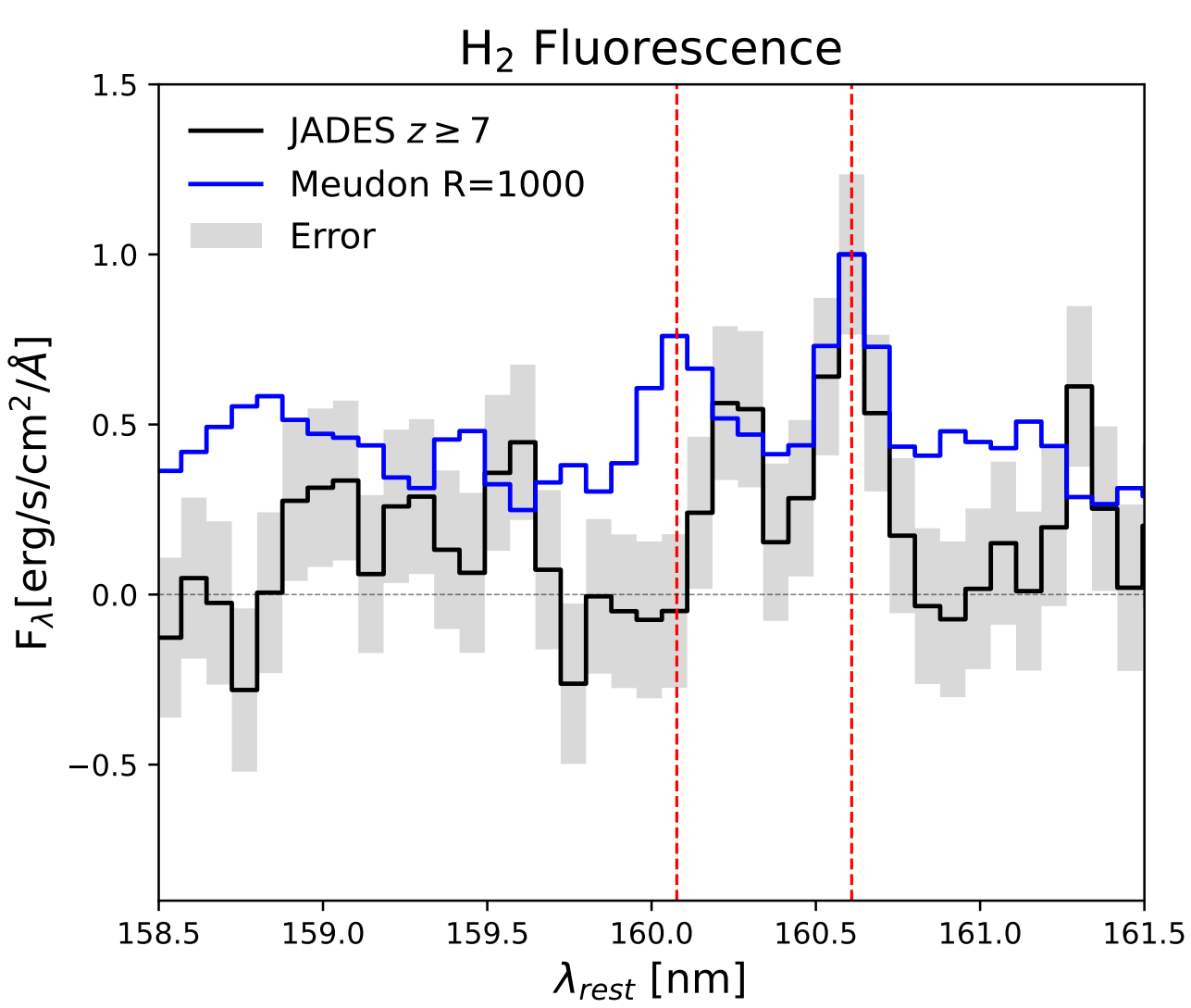

**Figure 10.** Comparison of the inverse-variance stacked $H_2$ fluorescent emission feature with our high-redshift Meudon model, which has been shifted by 0.18 nm or 340 km $s^{-1}$ (blue line). This displays the unique profile of the potential detection, which could indicate the presence of molecular outflows.

**Table 3**
Tentatively Detected Emission Lines, Their Measured Emission Wavelength, and Their SNR for the Inverse-variance Weighted Stacking of JADES Galaxies With $z \geqslant 7$

| Emission Line | $\lambda_{\rm obs,inv}$ (Å) | $\rm SNR_{inv}$ |
|---|---|---|
| C IV $\lambda\lambda$1548,1550 | 1544.49 | 4.28 |
| $H_2$ $\lambda\lambda$1603,1608 | 1602.23, 1606.09 | 3.68, 5.39 |
| He II $\lambda$1640 | 1640.74 | 4.38 |
| O III] $\lambda\lambda$1660,1666 | 1662.30, 1666.15 | 3.36,5.18 |
| C III] $\lambda\lambda$1907,1909 | 1906.39, 1909.47 | 1.66, 1.25 |

and SNR = 4.38, respectively. The most prominent feature of the $H_2$ fluorescent spectrum (160.61 nm) has a comparable SNR (SNR ∼ 5.38), while the next brightest $H_2$ feature is seen at 160.35 nm with SNR = 3.68. The SNR of the second feature is higher than that of the other stack presented. However, because of the discrepancy between the two methods, follow-up observations are still important and may allow us to detect these features with greater certainty.

## ORCID iDs

Madisen Johnson https://orcid.org/0009-0004-0577-2496
Blakesley Burkhart https://orcid.org/0000-0001-5817-5944
Francesco D'Eugenio https://orcid.org/0000-0003-2388-8172
Sandro Tacchella https://orcid.org/0000-0002-8224-4505
Roberto Maiolino https://orcid.org/0000-0002-4985-3819
Shmuel Bialy https://orcid.org/0000-0002-0404-003X
Jacques Le Bourlot https://orcid.org/0000-0003-3920-8063
Evelyne Roueff https://orcid.org/0000-0002-4949-8562
Franck Le Petit https://orcid.org/0000-0001-8738-6724
Emeric Bron https://orcid.org/0000-0003-1532-7818
Hervé Abgrall https://orcid.org/0000-0002-7099-8887
Erica Nelson https://orcid.org/0000-0002-7524-374X
Shyam Menon https://orcid.org/0000-0001-5944-291X
Matthew E. Orr https://orcid.org/0000-0003-1053-3081